\documentclass[runningheads]{llncs}
\usepackage[T1]{fontenc}
\usepackage{textcomp}
\usepackage{xcolor}

\usepackage{csquotes}
\usepackage{listings}
\usepackage{amsmath}
\usepackage{adjustbox}
\usepackage{graphicx}
\usepackage{algorithm}
\usepackage{algpseudocode}
\usepackage{longtable}
\usepackage{enumitem}
\usepackage{url}
\usepackage{wrapfig}
\usepackage{booktabs,tabularx,multirow}

\newcolumntype{Y}{>{\centering\arraybackslash}X}

\usepackage{acronym}
\usepackage{subcaption}
\newcommand{\customBold}[1]{{\fontseries{b}\selectfont#1}}

\begin{document}

\title{FIA Framework: A Large Language Model-Based Approach for Credit Card Fraud Investigations}

\titlerunning{A Framework for Credit Card Fraud Investigations}

\author{Shaun Shuster\inst{1} \and
Eyal Zloof\inst{1} \and
Rami Puzis\inst{1}
\and
Asaf Shabtai\inst{1}}
\authorrunning{}

%
\institute{Ben-Gurion University of the Negev}


\maketitle

\begin{abstract}
    Credit card fraud mitigation plays a significant role in modern society. 
    While fraud detection systems are essential, they often struggle to keep pace with the constantly evolving fraud techniques. 
    As a result, fraud investigation is an important complementary process required for continuously improving detection models, identifying emerging fraud patterns, providing case explanations of to stakeholders, 
    and maintaining customers' trust.
    However, fraud analysts are overwhelmed with an enormous number of alerts generated by credit card transaction monitoring systems.
    Each alert investigation requires careful attention, domain expertise, and thorough documentation of the investigation outcomes, leading to alert fatigue.
    To address this challenge, we introduce the first \emph{Fraud Investigation Assistant (FIA)} framework, which employs multi-modal large language models (LLMs) to automate key steps of credit card fraud investigation and generate explanatory reports.
    FIA leverages the reasoning, code execution, and vision capabilities of LLMs to collect relevant and logically consistent evidence while maintaining relatively short investigation trajectories. 
    Experiments with the Sparkov and CCTD datasets show that FIA gradually improves the F1 score while investigating borderline cases, reaching 8\%+ improvement after only 1,500 additional investigations.     
    These results suggest that LLM-based agents can assist with automating substantial parts of the fraud investigation process and may be particularly useful for resolving ambiguous alerts. 
    
    \vspace{0.5cm}
    \url{https://github.com/iamfriendly690/FIA-FRAMEWORK}

\keywords{Credit Card Fraud, Automated Fraud Investigations, Evaluation of Investigations, LLM}

\end{abstract}

\section{\label{sec:intro}Introduction}

    \textbf{Overview and motivation. }
    Credit card fraud refers to the unauthorized use of funds through credit or debit cards~\cite{bhatla2003understanding}.     
    Various techniques are used to steal the information needed to perform illegitimate online card-not-present (CNP) transactions, as well as physical transactions. 
    These methods include skimming, phishing, malware, and man-in-the-middle attacks~\cite{irvin2024identity,barker2008credit,peretti2008data}. 
    Given that credit card fraud losses exceeded 12.5 billion US dollars globally in 2024 alone~\cite{FTC2025fraudloss}, there is an urgent need to develop and implement effective strategies for fraud detection and investigation. 
        
    In most cases, banks and card networks evaluate transactions using rule-based systems and/or predictive models to identify known fraud patterns and assign risk scores to the transactions~\cite{wei2013effective}. 
    If the risk score for a transaction exceeds a certain threshold, an alert is flagged for manual investigation by a fraud analyst.
    Furthermore, real-world alert systems routinely select a subset of random transactions for manual review, regardless of their risk score.
    This random sampling is crucial for identifying new, unseen fraud patterns that existing detection models might miss, thereby providing the newly labeled data necessary to continuously retrain and update the models.
    Collectively, these investigations are essential for improving analysts' understanding regarding the nature of the transaction, providing case explanations to stakeholders (e.g., law enforcement), maintaining costumers' trust, and ensuring compliance with regulations.


    Credit card fraud investigations mainly involve gathering evidences.
    At the end of an investigation, a detailed report describing the investigation of the case is produced~\cite{nikkel2020fintech,raghavan2013digital,reith2002examination}. 
    Such reports are used by a company's senior management, law enforcement agencies, and legal teams to make informed decisions and take appropriate actions~\cite{financialcrimeacademyroles2024}.
    The results of an investigation are usually fed back to the fraud detection systems as new detection rules or data for retraining detection models~\cite{8038008}.

    Investigation of potential fraud cases requires careful attention, domain expertise, and thorough documentation of the investigation outcomes~\cite{personable2024transaction}. 
    The growing number of investigated fraud cases, together with the resource-intensive nature of individual investigations, can quickly exhaust human workforce resources. 
    This challenge underscores the need for automation to ensure scalability and operational efficiency.
    Today, artificial intelligence and large language models (LLMs) are used in a variety of tasks that require advanced analysis capabilities, such as root cause analysis~\cite{roy2024exploring,chen2024automatic}, cyber threat explanation~\cite{kaheh2023cyber}, and incident summarization within smart policing~\cite{sarzaeim2024framework}.

    To the best of our knowledge, existing work has focused primarily on fraud detection, with no current solutions addressing the automation of credit card fraud investigations.
    In this paper, we present the \emph{Fraud Investigation Assistant (FIA)} framework, that utilizes LLMs to support the automation of fraud alert investigations.
    Our goal is to investigate whether LLMs can analyze large volumes of transactions to resolve credit card fraud investigation cases.

\begin{figure}[t]
    \centering
    \includegraphics[width=0.82\linewidth]{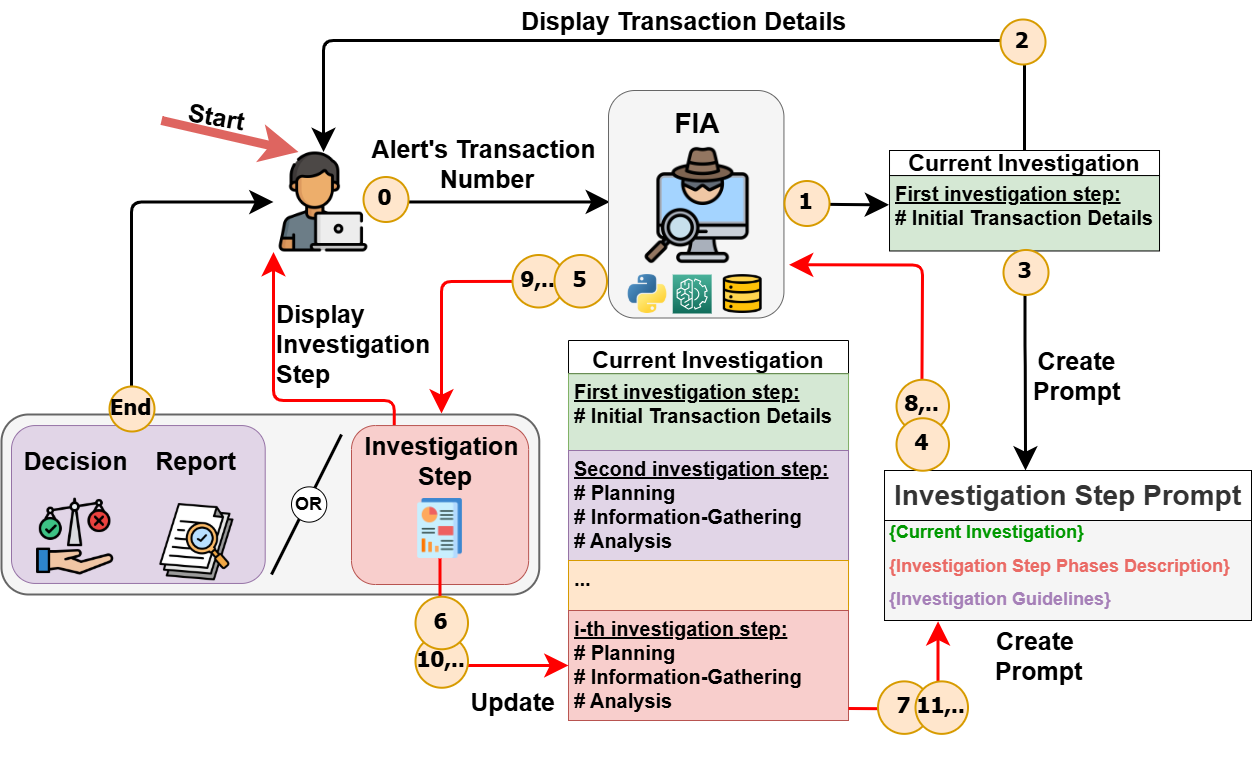}
    \caption{\label{fig:FIA_FRAMEWORK} Automated fraud investigation using FIA.
    }
    \end{figure}

    \textbf{\label{sec:FIA FRAMEWORK}High-level overview of FIA. } 
    The FIA framework comprises of LLM agents equipped with tools that operate through a series of investigative steps.
    Figure~\ref{fig:FIA_FRAMEWORK} illustrates the workflow of the proposed FIA framework. 
    The investigation process is initiated by an alert indicating a suspicious transaction, which is forwarded to the FIA.
    The framework is composed of iterative steps, where each step incorporates the current investigation, phase descriptions, and applicable guidelines. 
    These steps involve planning, data collection, visualization, and subsequent analysis.
    The investigation continues until sufficient evidence is gathered that either supports or refutes the fraudulent nature of the transaction in question.
    The framework then automatically generates a report summarizing the key evidence uncovered during the investigation.
    The generated report aims to reduce investigator workload, minimize alert fatigue, and mitigate human error by providing a concise summary of the major insights needed to resolve the case. 

    \textbf{Evaluation. }
    We evaluated the proposed FIA framework using the two well-known Sparkov~\cite{sparkov_apa} and CCTD~\cite{altman2019credit} datasets. 
    When using an LLM to classify based solely on the evidence in FIA-generated reports, the classifier achieved an F1 score of 98–99\%, with all metrics being higher for vision-enhanced investigations across both datasets, demonstrating that the framework extracts highly reliable insights for determining fraud.

    We evaluated FIA's utility when applied to the "uncertainty region" - the borderline cases where a state-of-the-art (SOTA) detection model is least confident.
    This targeted approach yielded a strong return on investment: by investigating just 1,500 alerts in detector's "uncertainty region", the FIA successfully increased its overall F1 score by 8.7–8.8\% across both datasets with around 0.5M transactions. 
    This shows the framework's value in efficiently resolving complex cases without completely depend on exhaustive, system-wide manual reviews.
    
    We identified that most investigations include only 20-40\% additional steps compared to the \emph{minimal trajectory}, which  consists of the essential steps identified in retroactively.
    In addition, we found that (1) 72\% of the evidence collected during investigations performed by the FIA had a high or very high impact on the suspicion level of the transaction in question, (2) all evidence collected were relevant, with no logical inconsistencies, and (3) 65-76\% of the collected evidence provided new information during the investigation.

    Furthermore, we observed that incorporating vision-based evidence significantly improved multiple aspects of the investigation. 
    Compared to investigations conducted without visual representations, vision-enabled investigations yielded a higher proportion of strongly aligned and novel insights (e.g., an increase of up to 16\% in “Strongly Agree” responses across logical alignment criteria).
    Additionally, vision-enhanced investigations produced more informative minimal trajectories with the price of increased token usage and slightly longer.
    

    \textbf{Contributions. }
     The contributions of this study are:
    \begin{itemize}[leftmargin=*]

    \item A fully automated framework for credit card fraud investigations that performs a complete investigation process and generates a detailed report.
    \item A method for deriving new evidence from large volumes of transaction data using vision multi-model LLM capabilities and dynamic code generation.
    \item An approach that converts an investigation step into structured LLM tasks.
    \item A new innovative and practical investigation evaluation measure for digital forensic investigations using LLM as a judge.
    \end{itemize}

\section{\label{sec:related}Related Work}

\subsection{Credit Card Fraud Detection}
\label{sec:Credit Card Fraud Detection}

Over the last decades, academia and industry have explored a variety of techniques to detect credit card fraud.
Most detection techniques revolve around rule-based systems~\cite{kumar2022rule,ramakalyani2012fraud}, 
machine learning (ML)~\cite{adaboost_ibm,feature_selection_ga_ibm}, 
and deep learning (DL)~\cite{xiang2023semi,lin2024fraudgt,harish2024leveraging,duan2025global}, and hybrid frameworks~\cite{Dhandore,Jemai2024}.
Less common approaches include symbolist, Bayesian, evolutionary, analogy-based, and graph-based approaches. 
See \cite{cherif2023credit} for a comprehensive survey. 

%


Predictive ML/DL models have notable shortcomings.
They struggle with generalizing to fraud techniques and types of new or underrepresented data in the training set.
Therefore, concept drift resulted by evolving fraud landscape is a major challenge. 
Additionally, opaque models often lack the explainability required to interpret and verify upon fraud alerts in high-stakes scenarios.

Recent studies employ large language models (LLMs) for financial fraud detection due to their ability to handle unstructured data~\cite{neha2024language,korkanti2024enhancing}.   
LLMs exhibit reasoning capabilities that are important for identifying novel fraud cases and increasing explainability. 
\cite{mori2025ai} shows that retrieval augmented generation can greatly increase credit card fraud detection accuracy compared to traditional ML, even with a simple architecture comprising a retriever, a generator, and an explainer agent.    
The author concludes that generative models provide the foundation for interpretability, compliance, and analyst support. 

Inspired by prior art, we expand the use of LLMs for the analysis of credit card fraud to a workflow that includes code generation and visual interpretation orchestrated to produce high-quality explanations. 
Explanations that highlight unequivocal evidence of committed fraud are required by the regulation before critical decisions can be made~\cite{uscode_fcba,cfpb_regz,eu_psd2,gao2021ai}.
Such evidence can only be collected through a deep investigation of forensic data related to each alleged fraud. 
In Section~\ref{subsec:evaluation_metrics}, we demonstrate how integrating the proposed FIA framework with SOTA detection models enhances the alert triaging process and improves overall detection performance.

\subsection{Investigation Challenges}
\label{sec:challenges}
Decisions made due to detected fraud cases must be backed by clear evidence. 
Thus, additional investigative frameworks~\cite{acfe2024fraud,caseiqultimateguide} are necessary to extract actionable insights, support evidence-based decision-making, provide case explanations to stakeholders (e.g., law enforcement), and maintain customer trust.

The main challenges faced by credit card fraud investigators are:

\textit{Complexity of Investigation. } 
Properly performed credit card fraud investigation process is complex, requiring attention, specialized expertise, and precision. 
It involves the retrieval of relevant data, discovering anomalous events, recognizing patterns, and performing various reasoning tasks. 

\textit{Alert Fatigue. } 
Like in cybersecurity, fraud analysts face many alerts raised by automated monitoring systems, leading to alert fatigue.

\textit{Documenting Evidence. }
It is crucial to meticulously document all evidence that supports the findings of an investigation and display a convincing reasoning chain leading to declaring a transaction fraudulent or legitimate~\cite{caseiqultimateguide}. 

In this paper, we attempt to automate most of the meticulous investigation process, focusing on identifying and documenting the most relevant evidence.

\subsection{Automated Investigation Using LLMs in Other Domains}
\label{sec:automatedInvestigations}
Recently, LLMs have been used to automate investigations in many domains, but not in credit card fraud; here we overview solutions in related domains.

Within the context of \emph{smart policing}, \cite{sarzaeim2024framework} suggest employing LLMs for crime classification and predictive analysis, producing a cohesive narrative.

The proposed framework performs well with a fixed task structure and investigation length. 
However, it may fail to capture evolving patterns or emerging anomalies not covered by a predefined schema.

\cite{roy2024exploring,chen2024automatic} use LLMs for cloud system failure root cause analysis. Both systems employ RAG to reason about past cases using retrieved logs, traces, and databases without retraining.

In \emph{digital forensics} domain \cite{isdfs2024} introduced a general framework for integrated digital investigations that operates based on natural language inputs. 
Their framework is built with a human-in-the-loop and employs a code generation agent for data retrieval. 

Automated LLM-based extraction of criminal information from digital evidence networks was proposed by \cite{zhoullm} to facilitate \emph{cybercrime investigations}. 
Their approach utilizes LLMs to learn patterns and relationships between forensic artifacts and enables the automatic construction of forensic intelligence graphs. 
LLMs can understand alerts from intrusion detection systems, analyze logs and threat feeds, and provide actionable, intuitive insights to human operators~\cite{kaheh2023cyber,jüttner2023chatids}.
These works utilize RAG to navigate the complex cybersecurity data and find evidence related to the investigated alerts.

To the best of our knowledge, FIA is the first framework that automates investigations of credit card fraud alerts. 
We rely on the work published in related domains and adopt their best methods, such as, using code generation for data analysis and transformation, mining diverse data sources with RAG, relying on reasoning models to perform multi-step investigations, and generating human-readable explanations that summarize the investigations.   
Furthermore, the FIA Framework contributes to the field of automated investigations by:  
(1) Highlighting diverse key performance indicators (KPIs) relevant to efficient investigations in addition to detection metrics employed by most other frameworks.  
(2) Analyzing the contribution of visual artifacts to the investigation performance. 



\section{Proposed Method\label{sec:method}}

    In this section, we first outline the various OpenAI GPT-5 mini agents and tools comprising the FIA, where the temperature was set to 1.0 for text-based agents and 0.7 for the vision agent. We then explain how these components are orchestrated using the Agents SDK~\cite{openaiagentssdk2026} to collaboratively perform an investigation.
    To ensure full reproducibility of the FIA workflow and the evaluation, all system instructions and specific prompts utilized by the FIA agents are publicly available in the GitHub repository.

    \subsection{Agents}
    \noindent\textbf{Orchestrator. }
    \label{subsec:orchestrator}
    The Orchestrator is a central agent responsible for coordinating the investigation step process. 
    It plans the strategy for each upcoming investigation step based on the current evidence and integrates new insights extracted from visual analyses.
    The planning process leverages the chain-of-thought (CoT) prompting technique~\cite{chen2023unleashing}, a widely used method to enhance reasoning in LLMs.

    Beyond strategic planning, the Orchestrator manages the interaction between agents: it instructs the Coding Agent to implement the investigative plan, provides new visual evidence to the Vision Agent for visual interpretation, and passes the collected evidence to the Report Generating Agent.

    In addition, the Orchestrator determines whether to continue or terminate the investigation, ending it only when no additional case-relevant direction is likely to change his current fraud assessment.
    Once this criteria had met, the orchestrator triggers the independent Detective Agent to assess the overall legitimacy of the case.
    Through these coordinated interactions, the Orchestrator ensures a coherent and iterative investigation workflow.

    \noindent\textbf{Coding Agent. }
    \label{subsec:codingAgent}
    The Coding Agent operates under the direction of the Orchestrator’s detailed plan for the current investigation step, specifying which data to extract from the transaction database, which analytical transformations to apply, and what specific patterns or anomalies to uncover to generate visualizations, allowing the Vision Agent to extract insights.
    In addition, the Coding Agent is aware of the data schemes it can operate on and without additional instruction can choose the right schemes to work with in the current step.
    
    The flexibility of the Coding Agent for dynamic analysis of each investigation is essential. 
    As fraudsters continually develop new evasion techniques, making a fixed set of actions insufficient, and execution of adaptive plan is mandatory.
    
    In order to perform its task, the Coding Agent uses a code execution tool to execute the code generated the Coding Agent in a sandboxed execution environment.
    This tool has access to transaction database for extracting relevant data.        
    If the agent's code fails, the code execution tool returns an error with detailed output error information, allowing the agent to iterate until it runs successfully. 
    
    \noindent\textbf{Vision Agent. }
    \label{subsec:visionagent}
    Using the descriptions provided by the Orchestrator with the visual artifacts generated by the Coding Agent, this agent's objective is to identify valuable trends, clusters, or anomalies within an artifact. 
    
    During the investigation step large volume of data is collected to be analyzed. 
    In the ablation study (see Section~\ref{sec:ImpactofVisionAgent}), we found the FIA struggled with analyzing the raw structured data without visualizations of the data.    
    
    \noindent\textbf{Report-Generating Agent. }
    \label{subsec:reportagent}
    Once the Orchestrator decides to stop the investigation steps, the agent generates a detailed report that summarizes the key evidence collected during the investigation.

    The generation process includes two steps:
    \begin{enumerate}
        \item First, the agent generates an unfiltered report that includes a description of all steps taken during the investigation and the evidence from each step.

        \item Then the agent takes the unfiltered report, and extracts the evidence it found to be the most important for determining the legitimacy of the suspicious transaction. This results in the report, which is presented to the analyst.
    \end{enumerate}

    \noindent\textbf{Detective Agent.}
    \label{subsec:detectiveagent}
    While not the primary focus, the generated report also enables a Detective Agent to determine whether the transaction is fraudulent based on the evidence provided in the report.

    \subsection{Fraud Investigation Assistant Workflow}
    \label{subsec:FIA}
    
\begin{figure*}[t]
            \centering
   \includegraphics[width=0.82\linewidth]{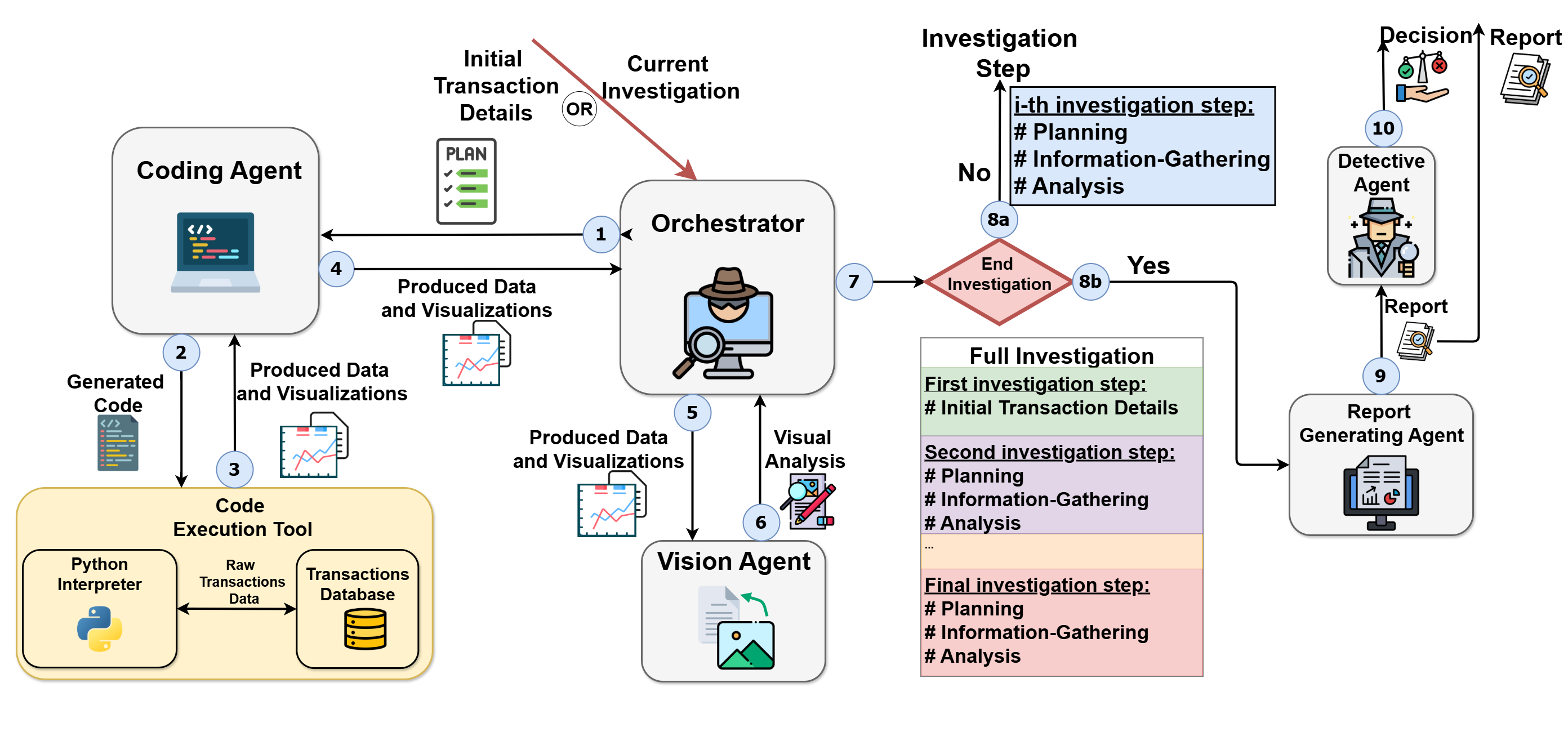}
            \caption{\label{fig:FIA_WORKFLOW}
            The FIA Workflow for performing a single investigative step. 
           }
        \end{figure*}

    The FIA framework operates through a sequence of investigation steps, each generated to iteratively build the investigation case. 
    This process is illustrated in Figure~\ref{fig:FIA_WORKFLOW}, and detailed in this section.
    Each step consists of a series of agent-driven operations that plan, extract, and analyze evidence related to the transaction.

    \noindent\textbf{Step 1: Initiating an investigation step. }
    At the start of each investigation step, the Orchestrator receives the current context, including the potential fraudulent transaction and insights from previous steps, and plans the next strategy.
    
    \noindent Note that in the first investigation step, the input provided to the Orchestrator includes only the information pertaining to the potential fraudulent transaction.

    \noindent\textbf{Step 2: Generating an investigation plan. }
    In this step, using the input received in Step 1, the Orchestrator is tasked with generating an investigation plan as defined in the planning prompt.
    
    \noindent For example, assume that in previous investigation steps, FIA identified that the transaction occurred 500 miles away from the cardholder’s home, raising suspicion.
    The plan provided in this step by the Orchestrator involves examining the geographic distribution of the cardholder’s transactions over the past week to determine whether distant transactions are typical or anomalous. 
    
     \noindent\textbf{Step 3: Generating code for the investigation plan. }
    In this step, the Orchestrator provides the plan and coding prompt, to the Coding Agent (action 1 in Figure~\ref{fig:FIA_WORKFLOW}).
    The Coding Agent generates code to execute the investigation plan.
    The code is generated in Python and executed via the code execution tool (action 2 in Figure~\ref{fig:FIA_WORKFLOW}) within an isolated virtual machine with access to the transactions database.
    Note that the Coding Agent's prompt directs it to extract textual and visual evidence, based on our hypothesis that LLM agents reason more effectively with visual aids from Vision Agent.
    However, since planning and code generation are handled autonomously in a fully agentic process, visual evidence may not be generated.
    The visualization and the produced data is then returned to the Orchestrator (actions 2-4).
    
    \noindent In our example, the Coding Agent generates code to retrieve all credit card transactions of the cardholder from the past week, maps their coordinates along with the cardholder’s home location and produces an image showing the map and transaction locations.  
    This visualization enables the Vision Agent to assess whether the recent activity was concentrated around the home location or included a suspicious outlier.
    

    \noindent\textbf{Step 4: Investigation step data augmentation. }
    If a new visual content is generated, the Orchestrator sends it to the Vision Agent (action 5 in Figure~\ref{fig:FIA_WORKFLOW}) along with the corresponding prompt.
    The agent is tasked with augmenting the investigation step data with new insights by identifying trends, clusters, or anomalies in the visualization. 
    The Vision Agent returns the results of the visual analysis to the Orchestrator (action 6 in Figure~\ref{fig:FIA_WORKFLOW}).
    
    \noindent In our example, the agent may identify two main geographic clusters: one near the cardholder’s home and one near the suspicious transaction.

    \noindent\textbf{Step 5: Investigation step analysis and conclusion. }
    In this step, the Orchestrator interprets the insights provided by the Vision Agent along with the investigation step data, the current investigation context (i.e., all information collected during previous investigation steps) and concludes this investigation step using the analysis prompt as a guideline.
    The Orchestrator can conclude that the collected evidence is not sufficient and that more analysis should be performed (i.e., starting from step 1 again).
    Otherwise, the Orchestrator can end the investigation and generate an investigation report.
    
    \noindent In our running example, the  Orchestrator concludes that there are two primary purchase locations, one of which includes the transaction in question. 
    This finding warrants deeper analysis of both locations to assess legitimacy, prompting the system to initiate another investigation step (action 8a).
    
    \noindent\textbf{Step 6: Generating an investigation report. }
    If the Orchestrator ends the investigation, it instructs the Report Generating Agent to produce a report and summarize all collected evidence (action 8b). 

    \noindent\textbf{Step 7: Case classification. }
    In the final step, the investigation report generated by the Report Generating Agent is passed to the Detective Agent to assess the legitimacy of the case, i.e., classificatory the case and a legitimate or fraudulent transaction (actions 9 and 10). 
        
\section{Experimental Evaluation}

\subsection{Research Questions}
To assess FIA's performance in fraud investigations, we aim to answer the following research questions:

\noindent\textbf{RQ1: How does the FIA's investigation process improve the underlying detection model, and what is the impact of targeting investigations within the model's uncertainty region?  
     }
     Fraud detection is not the main goal of FIA.
     Nevertheless, increasing the transaction classification measures is an important byproduct of the investigation process. 

\noindent\textbf{RQ2: Is the FIA framework capable of generating a report with high-quality evidence, without compromising the efficiency of the investigation process? }
     The report summarizes the investigation steps taken and the evidence gathered during the investigation.
     To evaluate the quality of the investigation we will evaluate the quality of the evidence shown in the generated report.
     We define a high quality evidence as an evidence that: 
     \textit{i)} The evidence has a significant impact on how strongly we perceive the transaction's legitimacy, as good evidence should influence our assessment of how legitimate the transaction appears.
     \textit{ii)} The evidence relevant to the investigation case.
     \textit{iii)} The evidence provides new knowledge.
     \textit{iv)} The evidence logically aligns with the rest of the investigation evidence.
     To answer this research question, we evaluate the quality of the evidence
     presented in the report using the evaluation measure described in Section~\ref{sec:EvidenceQualityScore}.
     The efficiency of the evidence extraction performed in the investigation process is measured by the number of investigation steps, the total number of tokens, and the minimal trajectory (Section~\ref{sec:investigation effectivness metrics}).

     
\noindent\textbf{RQ3: What is the impact of the vision agent the FIA framework? }
     We assess how including the vision agent in the FIA affects the results of the measures used to assess the FIA framework (RQ1) and (RQ2).

\subsection{Datasets}
    We evaluate our approach using two leading synthetic datasets: Sparkov~\cite{sparkov_apa} and CCTD~\cite{altman2019credit}, a high-quality dataset simulated from real-world data by IBM comprising multiple relational schemas. 
    Because the FIA framework models the human investigative process, it requires the non-obfuscated, human-readable features present in these benchmarks. 
    While some PII (e.g., names) can be masked, most features cannot be PCA-transformed or obfuscated in credit card fraud investigations without limiting investigative utility.
    Sparkov contains 1.85M card-not-present transactions (9,651 fraudulent), featuring comprehensive merchant, cardholder, and transaction details.
    To prevent LLM bias, we pre-processed Sparkov to remove the word 'fraud' from all merchant names.
    CCTD provides a similarly rich feature set across 24.3M legitimate and 29.7K fraudulent transactions. 
    After minimally preprocessing CCTD by casting numerical and binary columns to integers, we use the ground-truth labels from both datasets to assess the reliability of the reports generated by the FIA framework (RQ2).

    \subsection{Key Performance Indicators (KPIs)}
    \label{subsec:evaluation_metrics}
    To comprehensively evaluate the FIA framework, we establish measures across three key dimensions: fraud detection capability, investigation efficiency, and the quality of the generated evidence.
    
    \subsubsection{Fraud Detection Metrics}
    \label{sec:fraud detection metrics}
    Based on the results generated by the Detective Agent (Section~\ref{subsec:detectiveagent}), we measure the framework's ability to correctly classify fraudulent versus legitimate transactions.
    The agent's decisions are evaluated against actual transaction labels using the standard metrics most commonly employed in this domain: precision, recall, and F1-score~\cite{cherif2023credit}.
    By measuring this performance, we evaluate both the performance of the Detective Agent and the conclusiveness of the evidence provided in the investigation report.

    Investigating every single transaction or generated alert is impractical due to resource limitations. Therefore, to evaluate the true utility of the FIA, we apply these detection metrics not just globally, but strategically to the "uncertainty region"---the borderline cases where the underlying baseline detection model is least confident. 
    Measuring performance specifically on these borderline alerts allows us to quantify how effectively the framework resolves ambiguity and improves overall accuracy when we cannot afford to investigate all alerts.

    \subsubsection{Investigation Efficiency Measures}
    \label{sec:investigation effectivness metrics}~
    
    \noindent\textbf{Number of Investigation Steps:} This measure simply counts the number of investigation steps. We aim for a low number of steps.
        
    \noindent\textbf{Number of Tokens:} Recent studies have tried to reduce costs by simplifying queries, which can decrease the number of input and output tokens~\cite{shekhar2024towards}, or by dynamically selecting between more affordable and expensive models~\cite{bang2023gptcache,ong2024routellm}. Minimizing the token count helps reducing costs and improving response times.
        
    To count the number the text tokens, we used the "o200k\_base" tokenizer~\cite{tiktoken_github_2024}.
        
    \noindent\textbf{Investigation Minimal Trajectory Ratio:} An investigative trajectory often includes dead ends caused by incorrect hypotheses.
    To measure efficiency, we calculate the minimal trajectory ratio: the proportion of steps that actively supported the final decision.
    Denote $SDS(I)$ as the set of supporting decision steps and $STEPS(I)$ as the total steps in the investigation $I$, we define:
    \[
    \text{Min-trajectory-ratio}(I) = \frac{\lvert SDS(I) \rvert}{\lvert STEPS(I)\rvert}
    \]
    We retroactively determine $SDS(I)$ using an independent LLM to flag steps that contributed to the final decision.
        
    \subsubsection{Evidence Quality Score}
    \label{sec:EvidenceQualityScore}
    In the FIA framework, the report plays a major role, as it summarizes the investigation steps taken and the evidence gathered during the investigation. 
    Therefore, we would like to use the generated report for the evaluation of the framework. 
    However, since we want to evaluate the entire investigation process, there is a need to review all of the evidence gathered -- not just the evidence selected in the evidence filtering phase as described in Section~\ref{subsec:reportagent}. 
    Therefore, in our evaluation, we use the unfiltered report.
    Given the specialized nature of fraud analysis and the absence of an in-house domain expert, we adopted the established "LLM-as-a-judge" methodology.
    To rigorously mitigate circularity and model self-preference bias, the evaluator LLMs were assigned a distinct objective from the generation phase. Furthermore, we ensured robustness by averaging the evaluations of two leading non-OpenAI models: Claude 4.5 Opus and Gemini 3.1 Pro.
    We implemented this measure using a few-shot prompting technique to prompt an LLM to provide a rating on a five-point Likert scale for each piece of evidence on 4 different aspects as follows:

\noindent\textbf{Impact on fraud suspicion level:} 
            This evidence affects the level of fraud suspicion, either by increasing or decreasing it.

\noindent\textbf{Relevant to investigation case:} 
            This evidence was relevant to the investigation case.
            
\noindent\textbf{Providing new knowledge:} 
            This evidence either provided me with new insights or confirmed previously unverified information about the case.
            
\noindent\textbf{Logical alignment:} 
            The evidence logically aligns with the rest of the report.

\subsection{Fraud Detection Using FIA (RQ1)}
\subsubsection{Overall Classification Improvement}
We assessed the extent to which the final report generated by FIA (step 6 in Section~\ref{subsec:FIA}) helps improve the fraud classification accuracy. 
The transactions were classified by the Detective Agent based on the evidence presented in the report. 
\begin{table}[t]
  \centering
  \caption{Performance comparison of fraud detection models across Sparkov and CCTD datasets. Baselines are evaluated on the full test sets, whereas the FIA framework is evaluated as an investigative escalation step on a balanced sample of 500 transactions.}
  \label{tab:performance_metrics}
  
  \begin{tabular}{l ccc ccc}
    \toprule
    \multirow{2}{*}{\textbf{Model}} 
      & \multicolumn{3}{c}{\textbf{Sparkov}} 
      & \multicolumn{3}{c}{\textbf{CCTD}} \\
    \cmidrule(lr){2-4} \cmidrule(lr){5-7}
      & \textbf{F1} & \textbf{Precision} & \textbf{Recall} 
      & \textbf{F1} & \textbf{Precision} & \textbf{Recall} \\
    \midrule
    \multicolumn{7}{c}{\textit{Baseline Predictive Models (Evaluated on Global Test Set)}} \\
    \midrule
    \cite{dhandoreenhancing}*** & 0.850 & 0.870 & 0.840 & --    & --    & --    \\
    \cite{feature_selection_ga_ibm}** & --    & --    & --    & \textbf{0.998} & \textbf{0.999} & \textbf{0.998} \\
    \cite{adaboost_ibm}** & --    & --    & --    & 0.990 & 0.983 & 0.981 \\
    \cite{harish2024leveraging}       & 0.610 & 0.920 & 0.420 & 0.780 & 0.878 & 0.707 \\
    \cite{lin2024fraudgt}             & 0.762 & 0.866 & 0.680 & 0.792 & 0.910 & 0.701 \\
    \midrule
    \multicolumn{7}{c}{\textit{FIA Framework (Evaluated on a Balanced Sample of 500 Transactions)*}} \\
    \midrule
    \textbf{FIA}                      & 0.972 & 0.968 & 0.976 & 0.978 & 0.972 & 0.984 \\
    \textbf{FIA+Vision}               & \textbf{0.980} & \textbf{0.976} & \textbf{0.984} & 0.990 & 0.988 & 0.992 \\
    \bottomrule
  \end{tabular}
  
  \vspace{1ex}
  \scriptsize
  \begin{minipage}{\columnwidth}
    * \textbf{Evaluation Scope:} Because the FIA models an intensive human investigation process, it is evaluated on a uniformly sampled, balanced subset of 500 transactions (250 fraudulent, 250 benign). These metrics reflect its resolution capability on this subset and are not directly comparable to the global baseline metrics.
  \vspace{1ex}
    ** The results reported in \cite{feature_selection_ga_ibm,adaboost_ibm} could not be independently replicated. Our analysis indicates that the high performance metrics were likely an artifact of data leakage (e.g., applying SMOTE prior to train/test splitting).
  \vspace{0.5ex}
    *** The results reported in \cite{dhandoreenhancing} could not be independently replicated, achieving F1 0.710, Precision 0.760, and Recall 0.670 under strict data separation.
  \end{minipage}
\end{table}

Table~\ref{tab:performance_metrics} shows the results for standard detection KPIs: F1, precision, and recall scores. To mirror real-world alert systems that routinely select a random subset of transactions for manual review to identify new, unseen fraud patterns missed by existing predictive models, we evaluated the FIA's investigation capabilities on a balanced sample of 500 transactions.
For the Sparkov dataset, the FIA (with and without the vision agent) significantly outperforms all baseline methods across all KPIs within this sampled subset, demonstrating its strong capacity to uncover complex fraudulent activity. 
On the CCTD dataset FIA has surpassed the best-performing method. 

When evaluating the detection rate of the Detective Agent in FIA, we observed that the evidence provided in the reports generated by FIA is reliable for determining whether a transaction is fraudulent.

\subsubsection{Optimizing Investigations via the Uncertainty Region}
Building upon the motivation that exhaustive investigations are impractical, we must allocate the FIA's investigative resources to maximize return on investment (ROI).
As introduced in Section~\ref{sec:fraud detection metrics}, we claim that the FIA is most effective at the uncertainty region of SOTA models struggle to make confident decisions.

To evaluate this targeted approach, we centered the uncertainty-region analysis around FraudGT~\cite{lin2024fraudgt}, the best-performing reproducible baseline on both datasets.
The technical execution consisted of three steps. 
First, we trained FraudGT following its public implementation.
Second, using a separate validation set, we determined the decision threshold that maximized the F1-score and used it to define the model's uncertainty region.
Third, we applied this fixed threshold to the test set, ranked test transactions by their proximity to it, and investigated the resulting borderline alerts with FIA.
For each investigated alert, the Detective Agent's final binary prediction (see Section~\ref{subsec:detectiveagent}) was used as the resolved label for that transaction. This setup allows us to measure how allocating FIA to the most ambiguous test-time alerts affects the downstream detector.

\begin{figure}[t]
    \centering
    \begin{subfigure}[b]{0.495\textwidth}
        \centering
        \includegraphics[width=\textwidth]{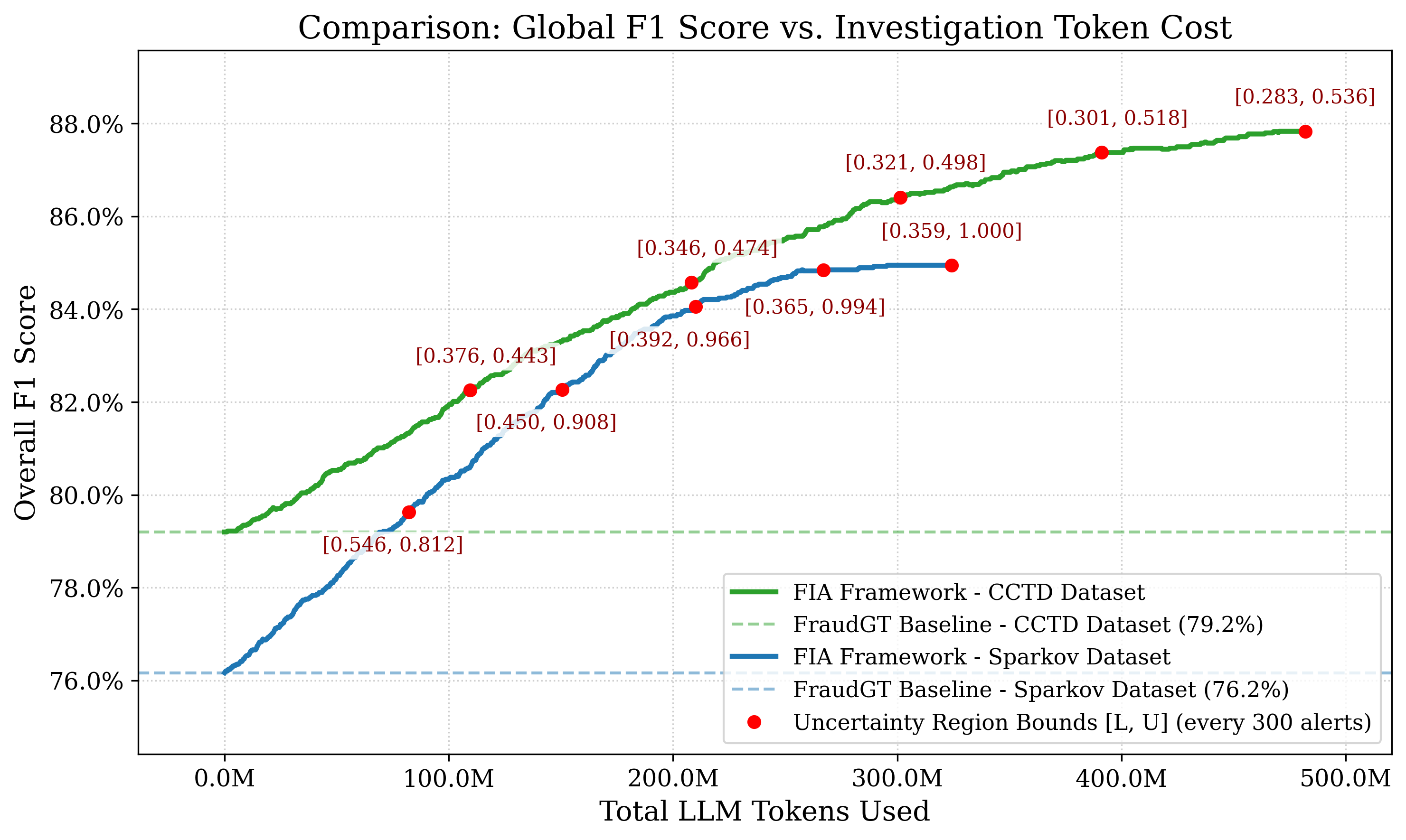}
        \caption{Global F1-score relative to total LLM token consumption.}
        \label{fig:f1_vs_tokens}
    \end{subfigure}
    \hfill 
    \begin{subfigure}[b]{0.495\textwidth}
        \centering
        \includegraphics[width=\textwidth]{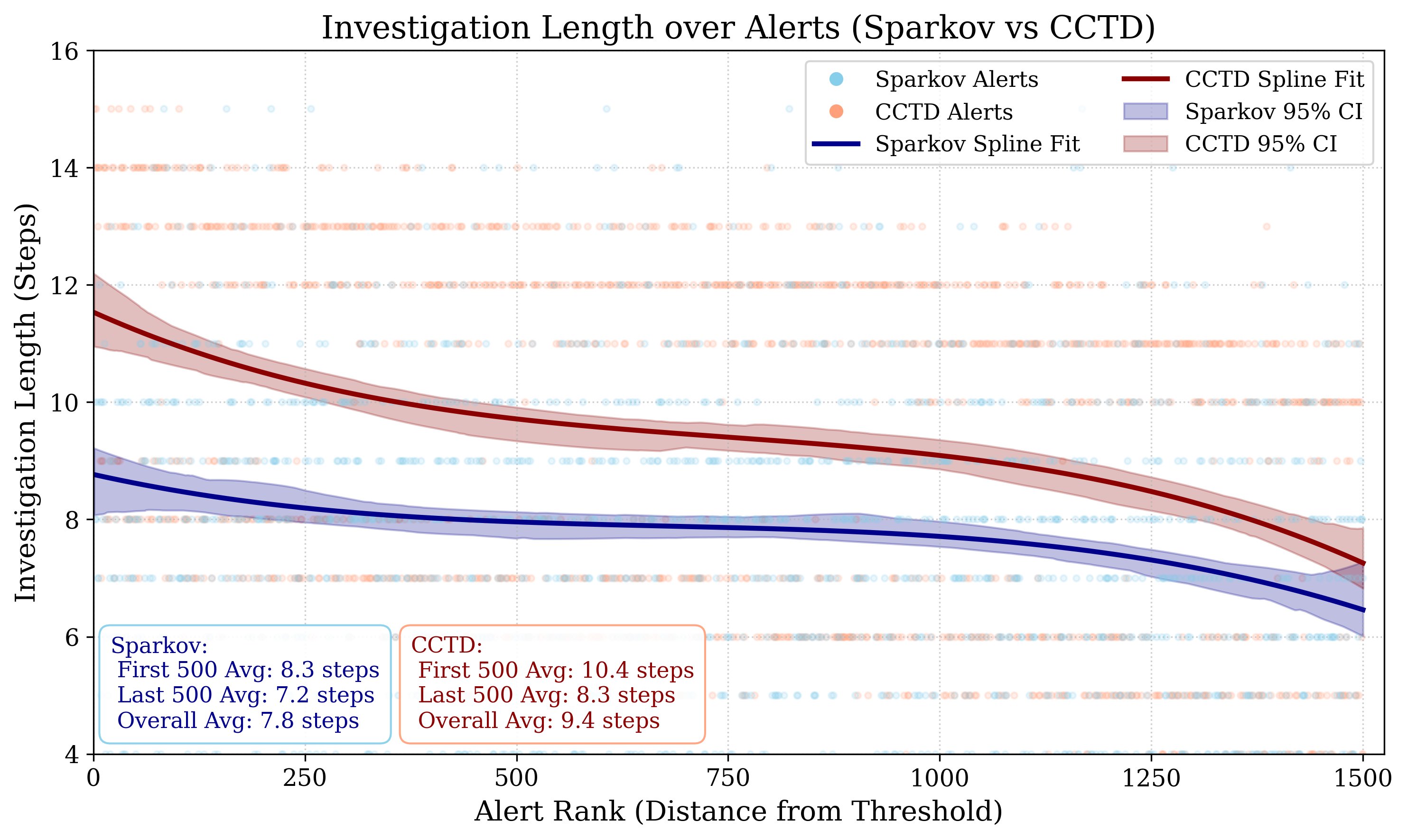}
        \caption{Correlation between investigation length and alert rank.}
        \label{fig:length_vs_difficulty}
    \end{subfigure}
    
    \label{fig:combined_f1_and_length}
\end{figure}

To quantify the framework's ROI, we evaluate the global F1-score against total LLM token consumption (Figure~\ref{fig:f1_vs_tokens}).
Strategically allocating resources to the uncertainty region yields substantial gains: investigating just 1,500 borderline alerts (merely $\sim$0.3\% of the $\sim$0.5M test transactions) increased the overall F1-score by 8.7--8.8\% across both datasets.


Furthermore, investigative effort correlates directly with model uncertainty (Figure~\ref{fig:length_vs_difficulty}). Alerts closest to the decision threshold require noticeably longer investigations: the 500 most uncertain alerts averaged 8.3 steps for Sparkov and 10.4 steps for CCTD, compared to just 7.2 and 8.3 steps for the 500 least ambiguous cases.
This confirms that highly uncertain cases are inherently more complex, justifying the deployment of the FIA framework's multi-step reasoning.

\begin{figure*}[t]
    \centering
    \includegraphics[width=\linewidth]{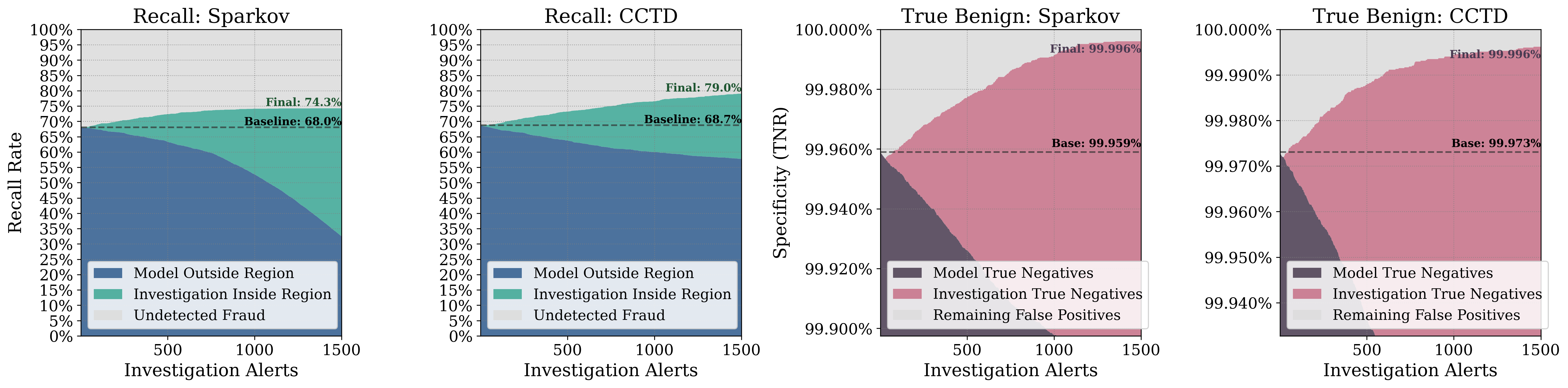}
    \caption{Stacked area plots depicting the cumulative improvement in Recall (left) and True Benign Rate (right) for alerts in the uncertainty region.}
    \label{fig:stacked_metrics}
\end{figure*}

To explain the F1-score improvement, Figure~\ref{fig:stacked_metrics} details the framework's impact on baseline error rates. 
Investigating borderline cases successfully recovers undetected fraud, boosting overall recall from 68.0\% to 74.3\% (Sparkov) and 68.7\% to 79.0\% (CCTD).
Simultaneously, it resolves false alarms by correctly identifying benign transactions, raising global specificity to 99.996\% across both datasets.
This dual capability recovering missed fraud while filtering false positives where the baseline model is most vulnerable - drives the substantial ROI observed earlier.

\subsection{Quality of Generated Evidence and Efficiency Assessment (RQ2)}
To address the second research question, we assessed investigations' efficiency and quality using the KPIs defined in Sections~\ref{sec:investigation effectivness metrics} and~\ref{sec:EvidenceQualityScore}, respectively.

\noindent\textbf{Investigation Efficiency Assessment.}
To assess the investigation efficiency, we analyzed the minimal trajectory ratio, input tokens, output tokens, and number of investigation steps. 
Results presented in table~\ref{tab:minimal_traj_ablation} indicate that investigation trajectories included 7 steps on average. 
72-83\% of the investigation steps were assessed as supporting the final decision (see Section~\ref{sec:investigation effectivness metrics}). 
Although the investigations contained mostly necessary steps, unnecessary steps were revealed that did not contribute significant evidence toward deciding the case.
Furthermore, the total number of tokens used was substantial, averaging above 205k tokens in a single investigation when utilizing the vision agent.

\noindent\textbf{Quality of Generated Evidence.}
Figure~\ref{fig:evidence quality comparison} presents the results of our evaluation of the quality of the evidence collected by the FIA on a 5-point Likert scale across four aspects: impact on fraud suspicion level, relevance to the investigated case, providing new knowledge, and logical alignment.
In the investigations we conducted, all of the evidence were found relevant to the investigation case and logically aligned with the rest of the investigation ("agree" or "strongly agree").
Some of this evidence, while relevant, may have contradicted the final decision made by the Detector Agent, as evident from 72-83\% minimal trajectory ratio.  

More than three-quarters of the evidence had high or very high impact on the level of fraud suspicion. 
Roughly the same number of evidence pieces were found as providing new knowledge about the investigated case. 


However, while the 'impact on fraud suspicious level' and the 'providing new knowledge' aspects were 71-72.14\% and 65-76\%, respectively, these results are raising some concern as they show that some part of the evidence evidence had questionable contribution to the investigations.


\begin{table*}[ht]
  \centering
  \caption{Comparison of the FIA framework under vision-enabled and non-vision settings across the Sparkov and CCTD datasets. 
    }
  \label{tab:vision_comparison}
  \label{tab:minimal_traj_ablation}
  \begin{tabularx}{\linewidth}{l l Y Y Y Y}
    \toprule
    \textbf{Dataset}
      & \textbf{Setting}
      & \textbf{Minimal Trajectory Ratio (\%)}
      & \textbf{Input Tokens (K)}
      & \textbf{Output Tokens (K)}
      & \textbf{Number of Investigation Steps}\\
    \midrule
    \multirow{2}{*}{Sparkov}
      & w/o Vision & 0.76 & 110  & 4.1 & 5.5\\
      & w/ Vision    & 0.80 & 205 & 5.05 & 6.0\\
    \addlinespace
    \midrule
    \multirow{2}{*}{CCTD}
      & w/o Vision  & 0.8 & 130  & 4.3 & 7.0\\
      & w/ Vision & 0.83 & 250  & 5.1 &  7.3\\
    \bottomrule
  \end{tabularx}
\end{table*}

These findings highlight the strengths of our FIA framework in generating structured and logical investigative steps while maintaining a high efficiency, which can be seen by, the high minimal trajectory.

\subsection{Impact of Vision Agent on the Investigations (RQ3)}
\label{sec:ImpactofVisionAgent}
To examine the contribution of the Vision Agent, we re-evaluated the same set of transactions used in the previous research question. 
This time, we excluded the Vision Agent and relied solely on the Augmented Data, allowing us to compare its impact on performance and investigation quality.

\begin{figure}[t]
    \centering
    \includegraphics[scale=0.6]{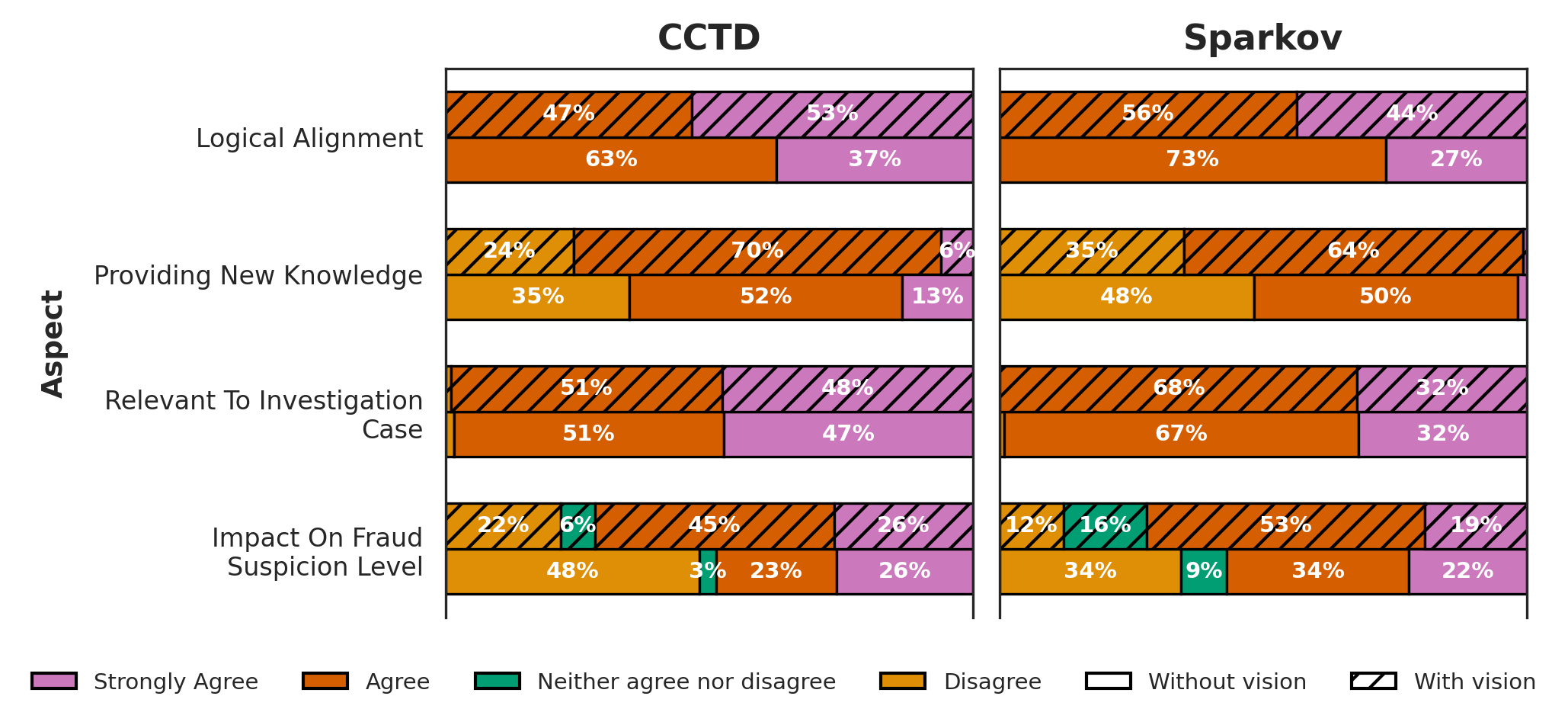}
    \caption{\label{fig:evidence quality comparison} Evidence quality comparison with vs. without the vision agent.}
\end{figure}

Figure~\ref{fig:evidence quality comparison} compares the quality of the between investigations conducted without the vision agent and those with it, each stacked bar represents the agreement level (strongly agree, agree, neither agree nor disagree, disagree) for different aspects of evidence quality: impact on fraud suspicion level, relevance to investigation case, providing new Knowledge, and logical alignment.

The results indicate that including the vision agent leads to higher levels of agreement in terms of evidence quality, particularly in the aspects of providing new knowledge and logical alignment, demonstrating the Vision agent's positive impact on our FIA framework, for both Sparkov and CCTD datasets.

Table~\ref{tab:minimal_traj_ablation} explores the trade-offs between token usage and trajectory efficiency under the conditions of "w/o vision" and "w/ vision".
The addition of vision capabilities enhances the minimal trajectory ratio, suggesting more efficient paths, but it also significantly increases both the input and output tokens.
This trade-off highlights that while the vision agent improves trajectory optimization, as seen in the higher minimal trajectory ratio, it requires more token usage. 
The impact on the number of steps remains moderate, with minimal variations observed between the two conditions. 
These findings underscore the benefits and costs associated with integrating the vision agent into the investigative process, emphasizing the need to balance efficiency gains with resource expenditures.

Table~\ref{tab:performance_metrics} shows an improvement in the detection rate performance when incorporating the vision agent. On the Sparkov dataset, the proposed framework with the vision agent achieved an F1 score of 0.9801, compared to 0.972 without vision capabilities. The CCTD dataset showed an even greater improvement, with the F1 score increasing from 0.978 to 0.99. These results highlight the effectiveness of incorporating the vision agent into the FIA framework.

\section{Conclusions}
In this paper, we introduce an innovative approach to automate credit card fraud investigations using LLMs.
The FIA framework leverages a combination of planning, information-gathering, and analysis phases to dynamically investigate suspicious transactions.
By integrating a variety of agents, the FIA framework automates critical steps in the investigation, and provides detailed and structured reports that enhance analysts' decision-making capabilities.

\noindent\textbf{Addressing the Challenges}
Our FIA framework tackles several core challenges inherent in credit card fraud investigations:
   \begin{enumerate}
        \item \customBold{Volume of Alerts}: Automating investigation steps, FIA reduces the manual workload of fraud analysts, allowing them to focus on more complex cases and reducing alert fatigue associated with handling a large volume of alerts.

        \item \customBold{Complexity of Investigation}: The FIA framework's systematic approach addresses the complexities involved in fraud investigations. Each step is planned, executed, and analyzed, ensuring a thorough examination of multiple aspects of each transaction it investigates.

        \item \customBold{Documenting Evidence}: The FIA framework automates the creation of detailed and well-structured reports. These reports offer a summary of the key evidence gathered during the investigation, such summary can help the analyst avoid incorrect decisions. 
        Making the report especially valuable for senior management, law enforcement agencies, and legal teams which relies on the significant evidence and the decision. We evaluate the report based on impact on fraud suspicious, relevance, logical alignment, and new information of its evidences.
    \end{enumerate}

By addressing these challenges, the FIA framework  enhances the efficiency, accuracy, and transparency of credit card fraud investigations, Leveraging the power of LLMs, the FIA framework not only reduces the burden on fraud analysts but also improves the overall efficiency of fraud investigation efforts.

For future work, we aim to study how incorporating continuous learning could impact the performance of the FIA framework.

This is intriguing because, although the framework generates reliable and efficient automatic investigations with high-quality evidence, it differs from a human fraud analyst, as human analyst learns about current trends in credit card fraud through experience, but the FIA framework is not aware of emerging fraud patterns since it does not learn from new investigations.

\bibliographystyle{splncs04}
\bibliography{references}

\onecolumn

\appendix

\section{APPENDIX}

\subsection{Dataset Memorization Check}
Inspired by~\cite{bordt2024elephants}, we evaluate whether GPT-5 mini memorizes our dataset using random feature completion on 1000 fraudulent and 1000 legitimate transactions.

\subsubsection{Random Feature Completion}
In this experiment We hid three columns per transaction (2000 × 3 = 6000 predictions) and task the model with recovering them. For features not inferable via domain knowledge, accuracy was 0, so no feature was correctly identified.
The following features were not included in this test, as they could be inferred using domain knowledge:
\begin{itemize}
    \item city - Using the zip or the state.
    \item state - Using the city or the zip.
    \item zip - Using the city.
    \item gender - Using the name.
\end{itemize}


\end{document}